# Measuring the Dynamic Impact of High-Speed Railways on Urban Interactions in China

Junfang Gong[1], Shengwen Li[1], Xinyue Ye[2], Qiong Peng[3]

[1]School of Geography and Information Engineering, China University of Geosciences, Wuhan, China

[2]Department of Landscape Architecture and Urban Planning, Texas A&M University, USA.

[3] Department of Urban Studies and Planning and National Center for Smart Growth, University of Maryland, College Park, USA

{ jfgong,swli}@cug.edu.cn, xinyue.ye@tamu.edu (corresponding author), xqpeng@umd.edu

Abstract: High-speed rail (HSR) has become an important mode of inter-city transportation between large cities. Inter-city interaction facilitated by HSR tends to play a more prominent role in promoting urban and regional economic integration and development. Quantifying the impact of HSR's interaction on cities and people is therefore crucial for long-term urban and regional development planning and policy making. We develop an evaluation framework using toponym information from social media as a proxy to estimate the dynamics of such interactions. This paper adopts two types of spatial information: toponyms from social media posts, and the geographical location information embedded in social media posts. The framework highlights the asymmetric nature of social interaction among cities, and proposes a series of metrics to quantify such impact from multiple perspectives – including interaction strength, spatial decay, and channel effect. The results show that HSRs not only greatly expand the uneven distribution of inter-city connections, but also significantly reshape the interactions that occur along HSR routes through the channel effect.

## 1. Introduction

The effects of regional rail investments is a longstanding topic of interest in the fields of regional/urban planning, real estate, housing, tourism, transportation, and economic geography because regional rail is a popular public investment used to promoting the economic agglomeration level of the regional economy, , mobility of skilled labor-force, and reduce greenhouse gas emissions (Chen et al., 2019; Pagliara & Mauriello, 2020; Yu et al. 2020;Shao et al.,2017). Researchers examine regional rail benefits from various perspectives, for example, a conventional perspective of investigating the effects is by comparing land or housing values along the rail corridor before and after the start of the rail service. However, fewer studies investigate the regional rail effects in terms of urban interaction, although promoting urban interaction is a primary objective for regional rail investment (Guo et al., 2020). Examinations of urban interaction resulting from regional rail project present important implications for benefit-cost analysis of regional rail investment and planning.

Furthermore, scholars of regional science, urban studies, and geography have long been interested in the interactions between cities and regions as they convey the spatial structure of a region (Zhen et al., 2019). The urban interaction involves the movement of people and cargo between places. The urban interaction also can refer to virtual connections, such as information communication between places (Ye & Wei, 2019). Conventionally, survey data were used to measure urban interaction, such as volumes of passengers between two cities (Xiao et al. 2013), migration flows (Flowerdew and Lovett 1988), trade flow (Hesse 2010), and telecommunications (Guldmann 1999). Compared with survey data, social media data offer a high temporal resolution, and are accessible at a large size with no or small cost.

High-speed rail (HSR), a type of regional trail commuting between cities and running much faster than traditional rail, significantly impacts urban systems , regional development, and the level of urban spatial interaction (Zhang et al., 2019). Due to their high speed and large capacity, it has largely become substitutes for domestic air transportation between cities (Chen, 2017, Gao et al., 2019). The development of high-speed rail (HSR) is a key urban infrastructure investment and economic development policy across a couple of countries (Wang et al., 2020b), especially for developing countries. HSRs are reshaping the maps of the spatial interaction, and reconstructing its network structure of cities (Teqi et al., 2005). These inter-city interaction and relationship brought by HSR tends to play a more prominent role in promoting urban integration and development in areas with more developed economies (Guo et al., 2020), including knowledge-intensive economy (Wang et al., 2020b). HSR network, the reduction in travel time has greatly increased the mobility of people across cities, thus affecting the spatial redistribution of the urban population in China, result in the phenomenon of urban shrinkage in China was further aggravated by the HSR (Deng et al., 2019). HSR Inter-regional commuting can be seen as a type of "temporary" migration (Guirao et al., 2018), give rise to a horizontal and polycentric city network and promote regional integration (Xu et al., 2019), and lead to a decrease in the disparity in population accessibility (Wang and Duan, 2018). Understanding and quantifying the impact of HSR's interaction on cities, therefore, is crucial for long-term urban and regional development policy and planning. Quantifying dynamic inter-city interaction brought by HSR is challenging. Commonly used indicators – such as per capita income or industrial output –fail to completely capture the effects of HSR. Besides, since these indicators describe the characteristics of individual cities, it is challenging to quantify the interaction between cities, especially if their mutual influences are uneven. Big spatio-temporal data extracted from social media not only contains rich spatial and temporal information, but it also contains information that details the interactions between cities (Han et al., 2015; Andris, 2016).

This paper develops a framework for measuring the bi-directional interaction between cities impacted by HSR by using a toponym from social media as a proxy. The contributions of this paper are summarized below:

1) Designing new metrics for HSR impacts by examining spatial attenuation effects and asymmetric interactions between cities.

2) Deriving a method to examine the channel effect of HSR, which refers to whether the two cities located at the endpoints of the HSR obtain far greater benefits than other cities through which the HSR passes.

3) Presenting a case study using social media data to demonstrate the applicability of the framework.

## 2. Related Work

## 2.1. Spatial Impact of HSR

Researchers have previously analyzed the line topology of HSR and studied the relationship and impact of HSR and urban spatial structure (Wang and Zhang, 2019, Zhang et al., 2019). Urban accessibility to HSR is evaluated using network analysis (Jiang et al., 2010). Using the gravity model to evaluate the HSR network, it was found that inter-city relatedness – as well as radiation and gravity models along the Beijing-Shanghai HSR – were consistent (Liao, 2015). The network centrality indicator was then used to investigate the spatial impact of HSR in urban networks (Cao et al., 2019, Wang et al., 2020a); it was found that, with the opening of the HSR network, overall connectivity between cities increased significantly (Jiao et al., 2017). These static studies focus on the spatial impact of the HSR line itself.

Previous research indicates that the development of HSR has greatly improved tourism, and affected labor migration and house values in small and medium cities situated along an HSR line (Chen and Haynes, 2015). There are other studies that take third-party data into consideration to measure the effect of HSR on everyday life and the urban economy. For example, survey data and regression models have been used to examine the effect of HSR on tourism (Guirao and Campa, 2015); labor and economic activity (Guirao et al., 2017, Li et al., 2016); and housing prices (Cheng et al., 2015, Diao et al., 2017), nighttime light data(Deng et al., 2020). The use of passenger flow data allowed for the spatial structure of 99 HSR cities in China to be analyzed, showing that the city network is multi-centered, particularly in the central and eastern cities (Yang et al., 2018). The impacts of high-speed rail development on airport-level traffic are examined by combing airport-level data, by considering not only the availability of air-HSR intermodal linkage between the airport and HSR station but also the position of the airport's city in the HSR network (Liu et al., 2019). HSR and inter-city coach timetable data have been utilized to evaluate city centrality and city-pair connectivity to compare hierarchical structures (Wang et al., 2020a). Nevertheless, researchers continue to face limitations in evaluating the impact of HSR. The spatial and temporal scales of economic and demographic data are limited by administrative districts, and some detailed data – such as the number of passengers and survey data of scenic spots – are difficult to obtain.

## 2.2. *Spatial Relatedness in the Big Data Era*

On the other hand, spatiotemporal interaction data is accessible via social media data, which is widely utilized in mapping disasters (Zheng et al., 2019), social activities (Tsou et al., 2013), public health (Hay et al., 2013), spatial planning (Massa and Campagna, 2014), and event detection (Sakaki et al., 2013). All of these studies illustrate the use of social media in the spatial social sciences. In examining HSR impact on tourism, for instance, check-in data collected revealed a high correlation between HSR and tourism. As such, this data can act as a proxy for real tourism arrivals (Liu and Shi, 2017). Existing studies have also focused on mining spatial and temporal distribution patterns, instead of measuring spatial interaction between geographic units.

Inter-city flows based on big data provide new perspectives for evaluating spatial interactions. Such connections can be derived using transportation and human travel data (Liu et al., 2010, Phithakkitnukoon et al., 2011), or telecommunications data (Gao et al., 2013), but these data are difficult to obtain due to privacy and security concerns. Social media account follower data have therefore been used to analyze topological features and to map the spatial distribution of inter-city social interactions in China (Li et al., 2013). Data of this type can also be challenging to work with, because user status updates visible on social media are rarely updated, making it difficult to measure the impact of HSR.

Toponyms (place names) from social media also provide a new type of data for mining spatial patterns and relationships (Meijers and Peris, 2019). Previously, this data has been used to examine how food environments influence food choices (Chen and Yang, 2014). It has also been used to address the connective strength and differences between geographical entities (Liu et al., 2014, Lin et al., 2019), and to simulate urban growth in metropolitan areas (Lin and Li, 2015). In addition, toponym data has been used to understand social media users' geographical awareness of U.S. cities by allowing surveyors to study how Twitter users exchanged and recognized toponyms from various U.S. cities (Han et al., 2015). Toponym co-occurrence results reveal the undirected interactions instead of directed interaction between two or more cities. Thus, the methods that rely on toponym co-occurrences cannot be directly used to quantify asymmetric interaction between cities.

Human behavior can also be reflected in the number of social media posts made by a person, or a community, or even as a whole, as well as the type of content posted. Social events – including emergencies, big events, holidays, and vacations – are common subjects for social media posts (Ben Lazreg et al., 2018, Xu et al., 2016, Ghani et al., 2019, Dolan et al., 2019). In this paper, we utilize toponymy data and geotag information from social media posts to model inter-city interactions and changes in these interactions resulting from HSR.

## 3. Data and Methods

Based on definitions used in existing studies (Han et al., 2015, Meijers and Peris, 2019), we use "awareness" to represent the basic interactions between cities implicit in social media data.

### 3.1. Data

This paper adopts two types of spatial information: toponym information from social media posts, and location information included in social media posts. A post with text that contains the toponym of city j, posted in city i represents an instance of interaction from city i to city j. We refer to this as awareness from city i to city j.

From a spatiotemporal perspective, inter-city social interaction extracted from social media can be expressed as a four-tuple (Toponym, location, Count, Time Duration). The "toponym" is the name of city j, the "location" is the name of city i, the "count" is the number of posts, and the "time duration" is the temporal slot for evaluation. A data example is shown in Table 1:

Table 1. Data example

| Record id | Toponym | Location | Count | Time Duration |
|---|---|---|---|---|
| 1 | Beijing | Beijing | 29899 | January 1, 2010 to January 31, 2010 |
| 2 | Beijing | Wuhan | 535 | January 1, 2010 to January 31, 2010 |
| … | … | … | … | … |
| 1001 | Beijing | Beijing | 220634 | October 1, 2010 to October 31, 2010 |
| 1002 | Beijing | Wuhan | 4282 | October 1, 2010 to October 31, 2010 |
| … | … | … | … | … |

## 3.2. A Framework

A greater value of awareness between two cities indicates a high level of interaction between the two cities. The suggested framework measures the strength of such relatedness from two dimensions. The first dimension describes the statistical perspective and the spatial perspective. The second dimension has two scales: global and channel effect. All inter-city interactions are considered and measured on the "global" scale, while the "channel" scale was developed to express characteristics between the terminal cities of an HSR.

The framework of relations between cities is defined from both the statistical dimension and the spatial dimension, as shown in Table 2:

Table 2. The Framework for assessing impact HSR

| Indices | Statistical dimension | Spatial dimension |
| --- | --- | --- |
| Global | Relatedness index | Spatial relatedness index |
| Channel effect | NA | Channel effect index |

In fact, the awareness between cities is directional. The large value of awareness of city A on city B on social media not only means that there is a stronger influence from city B to city A, but it also represents an equally strong dependence of city A on city B. As such, in-awareness is defined as awareness received from other cities, and out-awareness may be defined as awareness given to other cities. The indices in Table 2 consist of two sub-indices that evaluate both in-awareness and out-awareness, separately. The sub-indices of indices in Table 2 are listed in Table 3.

Table 3. The Indices List

| Main Indices | Sub-indices | Description |
| --- | --- | --- |
| City relatedness index | In-awareness index (IAI) | The sum of the in-awareness rate for a city |
|  | Out-awareness index (OAI) | The sum of the out-awareness rate for a city |
| Spatial relatedness index | Spatial in-awareness index (SIAI) | The sum of the spatial in-awareness rate for a city |
|  | Spatial out-awareness index (SOAI) | The sum of the spatial out-awareness rate for a city |
| Channel effect index (TS) | NA | A quantitative index of channel effect |

The various indicators in the framework of different time periods are examined, and the dynamic effects of HSR rail can be evaluated by observing the changes of the indices before and after the HSR opening.

In a specific period, the relatedness of N cities in social media may be represented by an N*N matrix from the four-tuples in Table 1. $C_{ij}$ is the awareness index, which is the number of posts by a poster located in city i, and whose text contains the name of city j, from city *i* to *j*. The larger $C_{ij}$ denotes the stronger awareness from city *i* to city *j*. Due to the differences between city *i* and city *j*, usually, $C_{ij}$ is not equal to $C_{ji}$.

### 3.2.1. Relatedness Index

Because $C_{ij}$ is represented by the number of posts between city *i* and city *j*, it is significantly connected with demographic factors (i.e. population, people composition, economy) of the two cities. In existing research, the populations of the two cities are employed to normalize the index. A drawback of this method is that the number of posts is not always significantly correlated with population. It is difficult to quantify the relationship between the number of posts and demographic factors. As shown in Figure 2 in section 4.1, the relationship between the number of posts and population varies in different cities.

Self-awareness is employed as the standard value to model inter-city relatedness, to consider the underlying population distribution. Both an awareness index and a dependence index are proposed in order to evaluate the regional relatedness, based on the regional asymmetry relatedness and the self-awareness.

To examine the total awareness of one city from all the other cites, the awareness index of city $i$ is calculated as:

$$IAI_i = \sum_{j=0}^{n} \frac{c_{ij}}{c_{jj}}, (i \neq j)$$

Where $n$ is the number of cities. The higher $IAI_i$ shows that the city i is more concerned by all the cities.

Not only does $C_{ij}$ represent the awareness from city $i$ to city $j$, but it can also be explained as an output between city $i$ and city $j$. For examining the dependence of a city to the totality of the cities, the output index of city i is calculated as:

$$OAI_i = \sum_{j=0}^{n} \frac{c_{ij}}{c_{ii}}, (i \neq j)$$

### 3.2.2. Spatial Relatedness Index

The Global Awareness Index (Han et al., 2015) estimates in-awareness based on spatial decay in social media. The mathematical expression for calculating what connects three of China's four first-tier cities – is given by:

$$GAI_i = \frac{\sum_{j=0}^{n} c_{ji} * G_{ij}}{P_i}, , i \neq j \tag{1}$$

Where $G_{ij}$ denotes normalized spatial distance, which is calculated by the actual spatial distance divided by the half length of the Earth's circumference; and $P_i$ is the population of the i-th city. Reasonably, the higher the GAI is, the more in-awareness there is, stemming from all the observed cities, and the greater the influence city i has on all the cities. In the $GAI$ model (Han et al., 2015), the population is used to normalize the value of $GAI$ to eliminate the impact of inter-city differences in size. For the same considerations as $IAI$ and $OAI$, this paper adopts local-awareness instead of population, to model the spatial awareness index. In the $SIAI$ model, $G_{ij}$ is adopted to denote the spatial weight between the i-th city and the j-th city. For a more general model, $G$ can be a spatial adjacency matrix, a distance matrix, or a reverse distance matrix.

Since spatial interaction is bidirectional, the index should be modeled separately in both directions. As a result, the $GAI$ is extended to the $SIAI$ index, and its calculation formula is as follows:

$$SIAI_i = \frac{\sum_{j=0}^{n} c_{ji} * w_{ij}}{c_{ii}}, i \neq j \tag{2}$$

Where $w_{ij}$ is the spatial weighted matrix, which is a variant of G. For n cities that are located along a HSR line, the $w_{ij}$ is denoted by:

$$w_{ij} = \frac{d_{ij}}{\max\{d_{ij} | i = 1..n, j = 1..n\}..} \tag{3}$$

For a city, the higher the *SIAI*, the stronger the influence there is in space. As with *SIAI*$_i$, the spatial out-awareness index is denoted by

$$SOAI_i = \frac{\sum_{j=0}^{m} C_{ij} * w_{ij}}{C_{ii}}, i \neq j \qquad (4)$$

Obviously, the higher *SOAI*$_i$ could represent that the i-th city has a stronger spatial out-awareness on all the other cities.

### 3.2.3. Channel effect

In general, cities at both ends of the HSR line tend to be important large cities. When an HSR line begins operations, the interaction between the two endpoint areas becomes greatly strengthened. How to quantitatively evaluate changes in interactions is a key concern for high-level decision makers and urban planners. To date, no related research on this topic has been observed in the existing literature.

According to economic theory, the channel effect refers to the behavior of majority shareholders, because they transfer capital and profits from the company to themselves (Carlson and Zmud, 1999). The channel effect suggests that the more resources that are gained from the given channel, the richer communication using that channel, and ultimately the richer the perception of the channel. In this study, the channel effect refers to whether the two cities located at the endpoints of the HSR obtain far greater benefits than other cities through which the HSR passes.

The channel effect is said to be reflected by changes in the spatial in-awareness index and the spatial out-awareness index. The spatial attention index here is adopted to estimate the channel effect because the channel effect mainly focuses on how to evaluate the unbalanced influence of cities in a particular geographical city. The channel effect can be estimated by

$$TS_{ij} = \frac{\frac{c_{ij}}{c_{jj}} * w_{ij}}{SIAI_j} = \frac{c_{ij} * w_{ij}}{c_{jj} * SIAI_j} \qquad (5)$$

Where the i-th city and the j-th city are the two cities in which the HSR line endpoints are located. The *TS*$_{ij}$ and *TS*$_{ji}$ are not equal due because their interactions are asymmetric. The two indices are combined into TS to check the overall channel effect of the high-speed rail line as follows:

$$TS = \frac{2TS_{ij} * TS_{ji}}{TS_{ij} + TS_{ij}} \qquad (6)$$

Under the randomization assumption, the expected value of TS is:

$$E(TS) = \frac{1}{n-1} \qquad (7)$$

When TS is greater than expected, it describes an indication that there is a channel effect in the research HSR line. The larger the value, the stronger the channel effect of the HSR line.

Similarly, a channel effect index can be derived from the perspective of economy as following:

$$TE_{ij} = \frac{GDP_j * w_{ij}}{\sum GDP_j * w_{ij}} \qquad (7)$$

Where $TE_{ij}$ is the channel effect from city *i* and city *j*, and $GDP_j$ is the GDP of city *j*. The *TE* index could be a reference for examining the *TS* index.

## 4. Case Study

### 4.1. Data

HSRs in China were first built in 2004 following an announcement by the Chinese government of a long-term plan for railway development. The Beijing-Tianjin high-speed railway that opened on August 1, 2004 was the first HSR line to reach 350 kilometers per hour. Since then, China built the largest HSR network in the world and has improved its existing railways over the course of a decade. As of the end of 2014, the total length of China's HSR exceeded 16,000 kilometers, accounting for more than 50% of the total length of the global HSR (Yanan, 2015).

China has built the Beijing-Shenzhen HSR line as shown in Figure 1 from 2005 to 2012. The line consists of three parts, the Wuhan-Guangzhou rail, the Guangzhou-Shenzhen rail, and Beijing-Wuhan began running on December 26, 2009, December 26, 2011, and December 26, 2012, respectively. The line connects many cities, and had been in operation for several years so far, accumulating a presence and influence on social media. Recognizing this, we use this line to conduct our research.

The Beijing-Shenzhen HSR line – which connects three of China's four first-tier cities – is the main transportation corridor from northern China to southern China. The HSR passes through 35 cities which are classified into four groups as shown in Table 4 by the size and the role of those cities:

Table 4. The cities located in Beijing-Shenzhen HSR line

| Group | Description | Number of cities |
|---|---|---|
| FC | First-tier city | 3 |
| SC | Provincial capital city | 4 |
| TC | Prefecture-level city | 17 |
| MC | Small city | 11 |

The first group (FC) is composed of Beijing, Guangzhou, and Shenzhen, which are three of four of China's first-tier cities. The second group (SC) consists of Shijiazhuang, Zhengzhou, Wuhan, and Changsha, which are provincial capital cities. The third group (TC) includes 17 prefecture-level cities. In addition, the MC group includes small cities and consists of county-level cities. Actually, every MC city is a part of prefecture-level city in the TC group, and so the cities in the MC group will be removed from the following experiments. The interaction between the first three groups (24 cities) was selected as the research focus for this study.

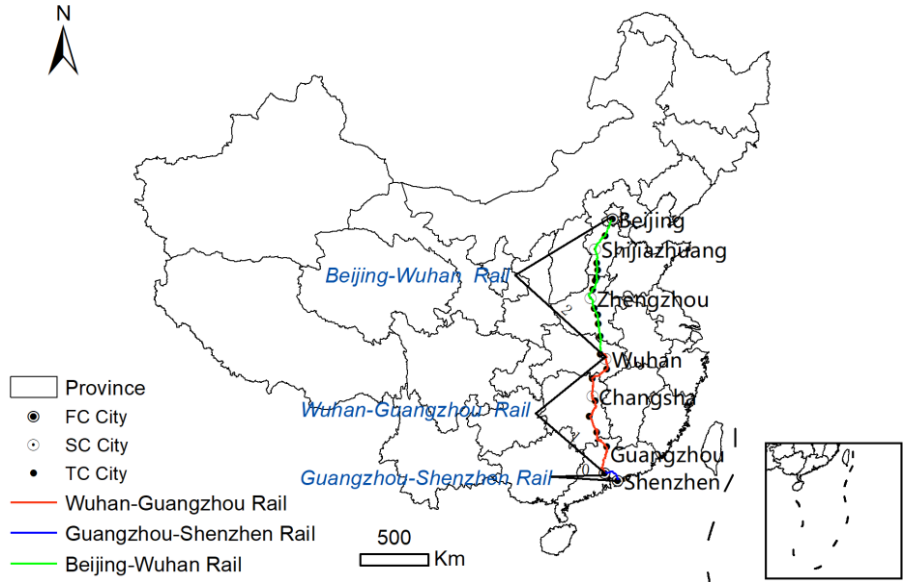

Figure 1. Cities and rails

Sina Weibo is a Chinese social media platform described as a cross between Twitter and Facebook platforms. Each post on Weibo is limited to 140 Chinese characters and some pictures. Weibo provides a web search query that can sort through and show posts – as well as provide the number of posts – based on filters such as post location, content, and time. A search URL example can be organized to "https://s.weibo.com/weibo/? q= 深圳®ion=custom:11:1000&typeall=1×cope=custom:2010-01-01-0:2010-01-31-23". Where "深圳" is the Chinese name of Shenzhen city; "custom:11:1000"is used to refine the posts located in Beijing cities; "typeall=1" means the return results will include all type of posts; "timescope=custom:2010-01-01-0:2010-01-31-23" means the return posts posted in January 2010.

A crawler was developed to simulate the process of post searches and the results were saved as a dataset. The dataset consists of a four-tuple (toponyms, location, count, time duration) as shown in Table 1. As such, the interaction between cities in each period forms an OD matrix.

Inter-city interaction $C_{ij}$ was recorded on a month granularity for the 24 research cities in the three groups. The dataset contained 576 records (24 * 24) per month. A total of 27,648 (24*24*48) records were accumulated over four years, and thus the datasets from one city to another city are comprised of 48 $C_{ij}$.

To examine the correlation between posts and HSR, the cities were classified into 2 groups. The first group cities named "JW" are located along the "Beijing-Wuhan-trail", and the second group cities named "WG" are located along the "Wuhan-Guangzhou-trail". The number of posts between the two groups and in each group were calculated and plotted in Figure 2. In December 2012, the "Beijing-Wuhan- trails" started running, so the "JW" and "WG" cities were directly connected by HSR. During this period it was observed that the number of posts from "JW" to "WG" increased rapidly. This is clear evidence that the posts in social media can reflect the changes in the relationship between cities.

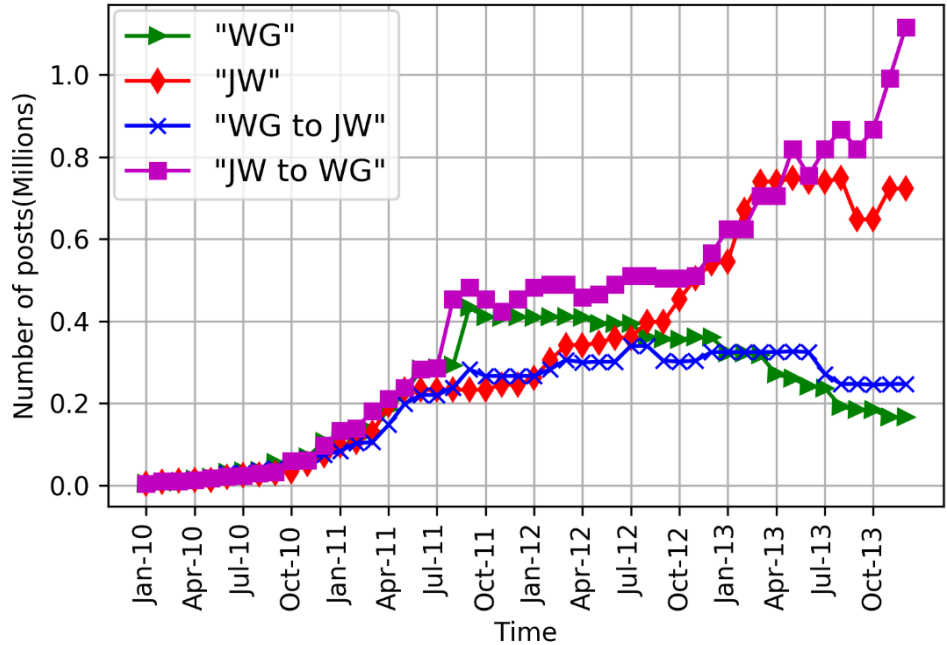

Figure 2. Number of posts of "JW" cities and "WG" cities

There are fluctuations in the number of posts per month as shown in Table 5, largely because posting behavior is closely related to human behavior, which carries a degree of uncertainty.

Table 5 The difference in post numbers in Beijing and Shenzhen over four years.

(a) The $C_{ij}$ (i=Shenzhen j=Beijing)

|      | max | min | avg | median | stdev | Total |
|------|-----|-----|-----|--------|-------|-------|
| 2010 | 11341 | 1101 | 4834.58 | 4640 | 2727.90 | 58015 |
| 2011 | 100966 | 9279 | 43905.83 | 36914 | 26918.02 | 526870 |
| 2012 | 134276 | 64953 | 92608.50 | 88473 | 20060.48 | 1111302 |
| 2013 | 197789 | 68046 | 104774.17 | 92480 | 45325.05 | 1257290 |

(b) The $C_{ij}$ (i=Beijing j=Shenzhen)

|      | max | min | avg | median | stdev | Total |
|------|-----|-----|-----|--------|-------|-------|
| 2010 | 13403 | 956 | 6392.00 | 7217 | 3655.86 | 76704 |
| 2011 | 434206 | 26806 | 99336.75 | 62891 | 106690.38 | 1192041 |
| 2012 | 407025 | 74232 | 173019.83 | 146402 | 86256.40 | 2076238 |
| 2013 | 522490 | 95883 | 271917.58 | 250533 | 114306.05 | 3263011 |

To reduce the uncertainty of the number of posts, records of every $C_{ij}$, were classified into four groups cataloged by the year. The median value calculated based on a calendar year was employed to represent the value of a year. The new data consisting of median values from each year contained 2,034 records for 24 cities over four years.

**4.2. Results**

### 4.2.1. Population and Inter-city Social Interaction

The population and four groups of $C_{ii}$ (the awareness from city $i$ to itself) were examined for these variations from 2010 to 2013. The data from the five groups are normalized by the maximum value of each group as shown in Figure 3. The cities in the figure are arranged according to total population from left to right.

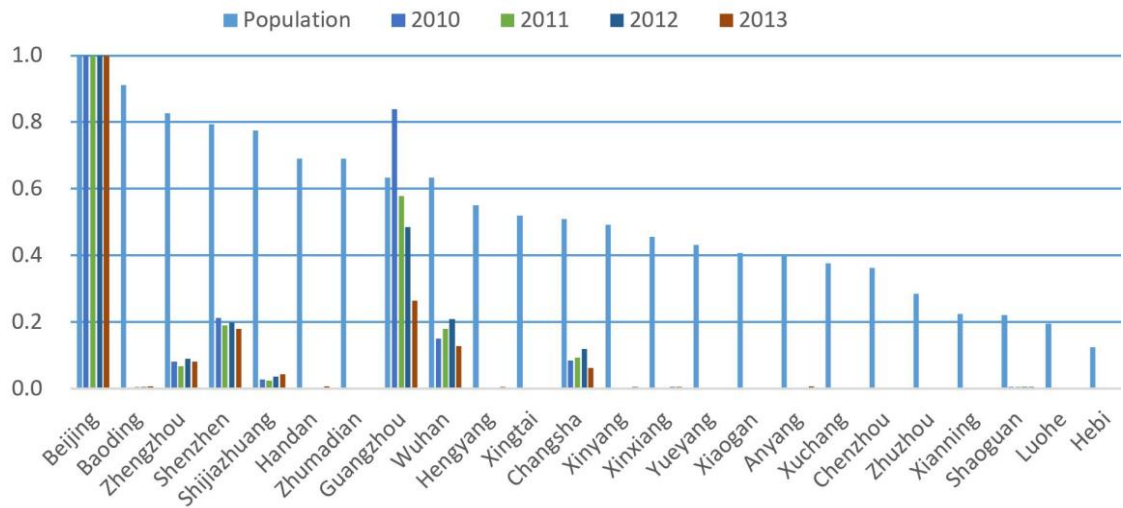

Figure 3. Difference between population and the number of local posts in four years from 24 cities

As shown in Figure 3, a noteworthy difference exists between the number of people and the value of local-awareness ($C_{ii}$). The value of $C_{ii}$ of three FC cities – Beijing, Guangzhou, and Shenzhen – is significantly higher than that of other cities. A Pearson correlation analysis is employed to examine the correlation between population and local-awareness of those cities. It reveals their differences, as the coefficients between population and $C_{ii}$ are 0.436, 0.486, 0.466, and 0.483 in the four years. It shows that population as a variable has some limitations in eliminating the impact of city size on human activities on social media.

The stability of $C_{ii}$ is further examined over time. As shown in Figure 3, the trend of $C_{ii}$ remains relatively constant over the four years. A Pearson correlation analysis was also conducted to verify the interaction of $C_{ii}$ in the four groups. The correlation coefficients between the groups are greater than 0.89, and their statistical significance was less than 0.001 (2-tailed). This means that, although social media user activity is uncertain, the $C_{ii}$ is statistically relatively stable.

The above analysis shows that $C_{ii}$ – or the local-awareness – likely reflects social media activities more accurately than the population for city i. In addition, the distribution of the basic indices is examined as well. The distributional pattern of $C_{ij}$ is shown in Figure 4. This distribution is long-tailed and consistent with the previous research (Scellato et al., 2010).

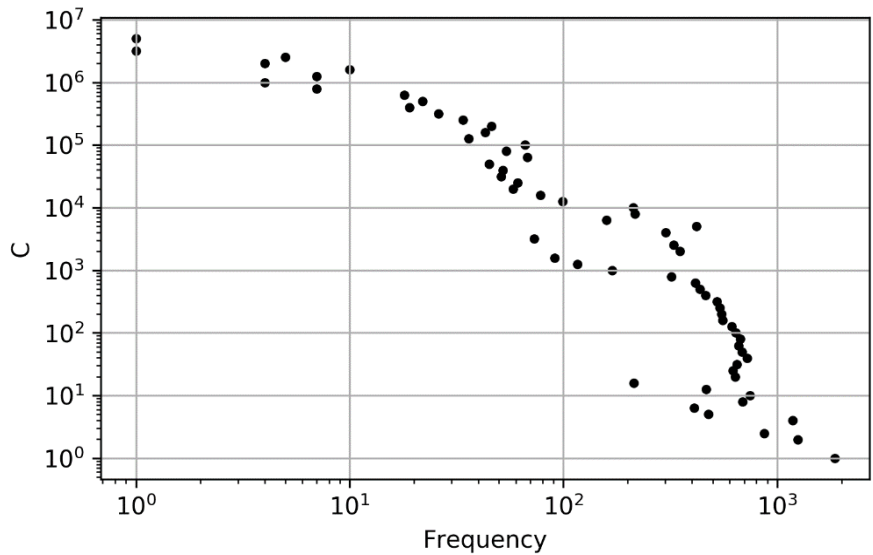

Figure 4. The distribution of $C_{ij}$

Finally, the significance of $C_{ii}$ is examined. It is seen that 66 of the total 2,034 records showed $C_{ij}>C_{jj}$, or where concern for other cities is greater than local-awareness. The number accounting for 3.24% of the total is statistically significant at the 0.05 level. The number of records of $C_{ij}>C_{ii}$ – which stands for cities with a greater in-awareness rather than a greater local-awareness – is a total of 14 of 2,034. The number accounting for 0.66% of the total record has a statistical significance at the 0.01 level. In other words, both inter-city in-awareness and inter-city out-awareness have significant spatial autocorrelation. In summary, it appears likely that the model based on $C_{ii}$ provides a better way to evaluate inter-city interaction than a model based on population.

### 4.2.2. Inter-city Relatedness

The inter-city in-awareness index and out-awareness index are calculated and shown in Table 6 from 2010 to 2013.

Table 6. In-awareness index and out-awareness index

|  | IAI | | | | OAI | | | |
| --- | --- | --- | --- | --- | --- | --- | --- | --- |
| Year | 2010 | 2011 | 2012 | 2013 | 2010 | 2011 | 2012 | 2013 |
| max | 16.690 | 14.490 | 13.070 | 10.170 | 2.900 | 4.240 | 2.360 | 2.620 |
| min | 0.010 | 0.020 | 0.020 | 0.020 | 0.240 | 0.370 | 0.260 | 0.320 |
| average | 1.632 | 1.390 | 1.121 | 0.920 | 1.462 | 1.375 | 0.949 | 0.831 |
| median | 0.110 | 0.180 | 0.170 | 0.140 | 1.530 | 1.200 | 0.670 | 0.645 |
| stdev | 3.599 | 3.075 | 2.715 | 2.129 | 0.760 | 0.887 | 0.644 | 0.510 |

Table 6 illustrates that the mean and standard variance of the inter-city in-awareness index decreases over time, and the median similarly decreases in each of the preceding three years. In general, the trends show that inter-city influence and city differences are decreasing. They also demonstrate that the inter-city dependence and its differences are lessening, and the mean, median, and standard of inter-city out-awareness also continue to decrease.

Although the total rate of each city's awareness from other cities is increasing (the conclusion in the last section according to the X index), the inter-city in-awareness index (IAI) and the trend of inter-city out-awareness index (OAI) are also on a downward trend. According to the definition of IAI and OAI, local-awareness has a higher growth than in-awareness.

Obviously, there are some gaps between the trends of IAI and OAI. These gaps are examined using the difference in the value of IAI and OAI in the same group in the first year (2010), and in the last year (2013) as shown in Figure 5.

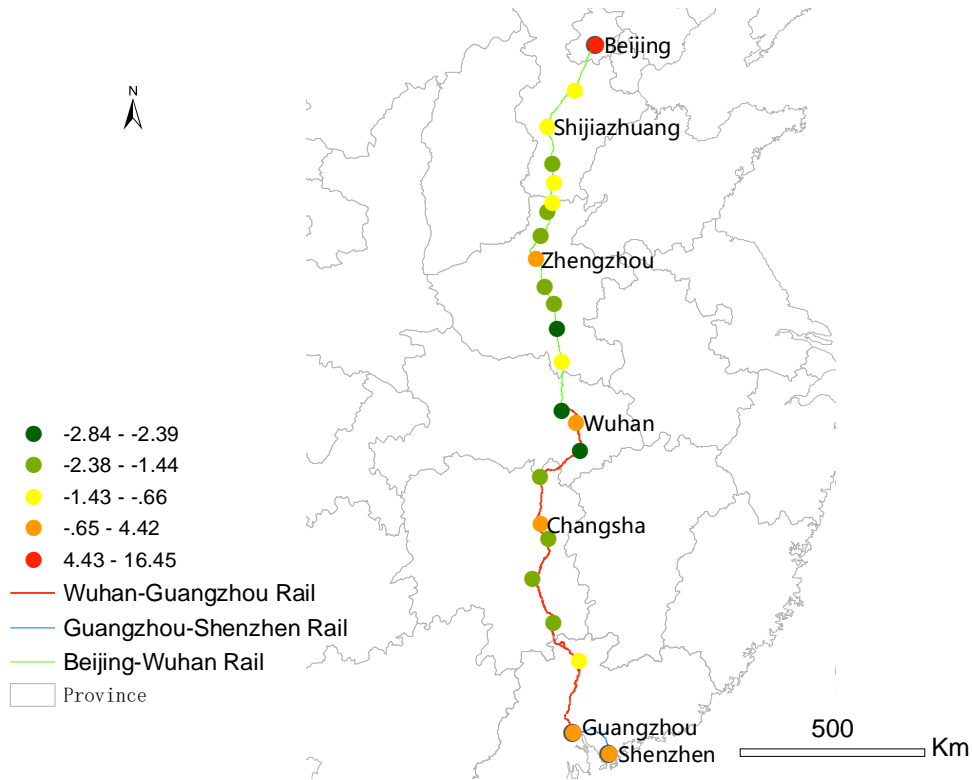

Figure 5(a) The difference (IAI- OAI) of cities between IAI and OAI in 2010

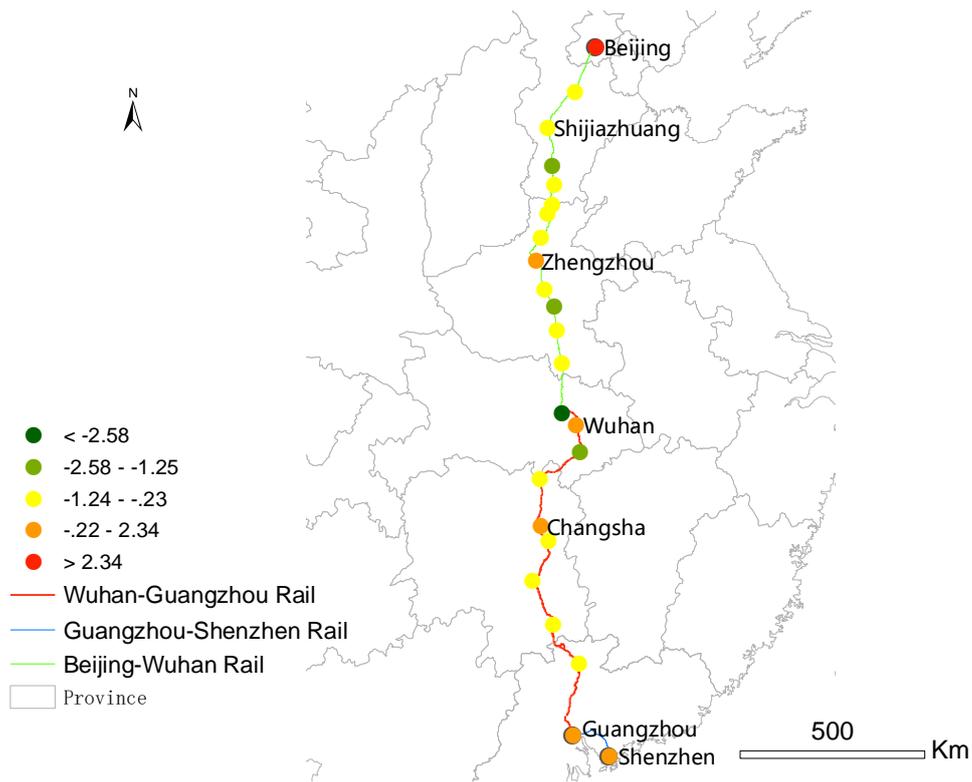

Figure 5(b)

The difference (IAI- OAI) of cities between IAI and OAI in 2013

As can be seen from Figure 5, in most cities, the value of out-awareness is greater than the value of in-awareness. To detail those differences, the ranked IAI and ranked OAI for seven central cities from 2010 to 2013 are listed in Table 7(a).

Table 7(a). Comparison IAI and OAI between in 7 central cites

| City | Type | Population | Rank (IAI) | | | | Rank(OAI) | | | |
|---|---|---|---|---|---|---|---|---|---|---|
| | | | 2010 | 2011 | 2012 | 2013 | 2010 | 2011 | 2012 | 2013 |
| Beijing | FC | 12.97 | 1 | 1 | 1 | 1 | 24 | 24 | 21 | 16 |
| Shenzhen | FC | 2.87 | 2 | 2 | 2 | 3 | 23 | 23 | 24 | 24 |
| Guangzhou | FC | 8.22 | 3 | 4 | 6 | 5 | 19 | 19 | 12 | 14 |
| Wuhan | SC | 8.21 | 4 | 5 | 4 | 6 | 21 | 21 | 18 | 17 |
| Changsha | SC | 6.60 | 5 | 3 | 3 | 2 | 22 | 22 | 15 | 21 |
| Zhengzhou | SC | 10.72 | 6 | 6 | 5 | 4 | 20 | 17 | 11 | 15 |
| Shijiazhuang | SC | 10.05 | 7 | 7 | 7 | 7 | 15 | 9 | 8 | 7 |

Table 7(b). Comparison IAI and OAI between the central cities and the TC cities

| Year | IAI | | OAI | |
|---|---|---|---|---|
| | Central city | TC city | Central city | TC city |
| 2010 | 5.3386 | 0.1053 | 0.5871 | 1.8218 |

| 2011 | 4.4486 | 0.1306 | 0.6957 | 1.6541 |
| 2012 | 3.5571 | 0.1176 | 0.6057 | 1.8218 |
| 2013 | 2.8886 | 0.1100 | 0.5886 | 0.9306 |

As can be seen from Table 7, the in-awareness index and out-awareness index are negatively correlated. The seven central cities (FC cities and SC cites) occupy the top seven spots in the rankings of the in-awareness index from 2010 to 2014, even though their out-awareness indices are very low.

### 4.2.3. Spatial In-awareness and Spatial Out-awareness

Spatial in-awareness and spatial out-awareness are used to evaluate inter-city interaction based on spatial decay theory. The values of these indices from 2010 to 2013 are shown in Figure 6.

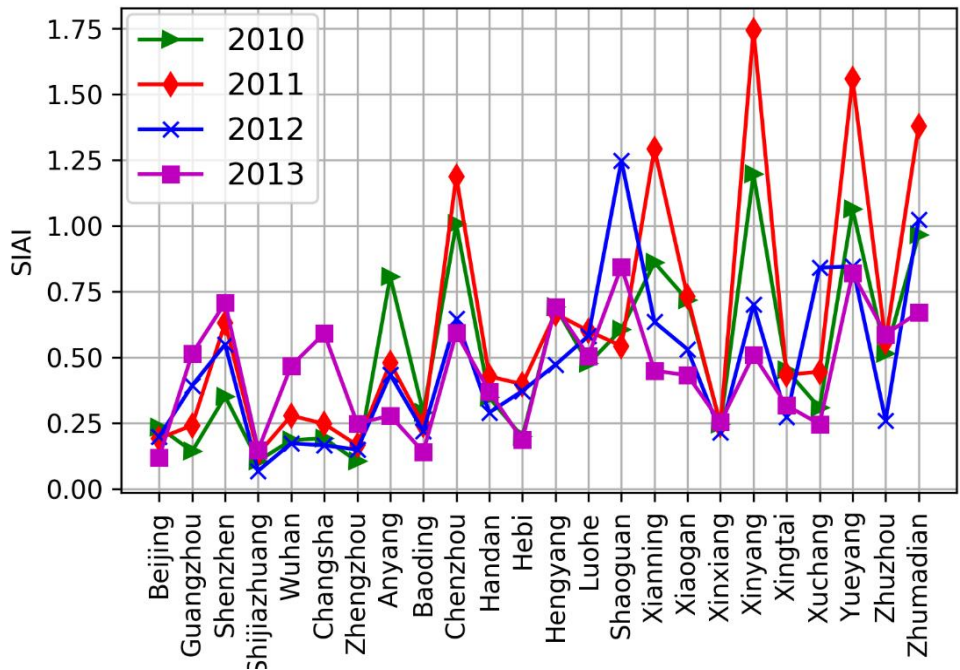

Figure 6(a). SIAI from 2010 to 2013

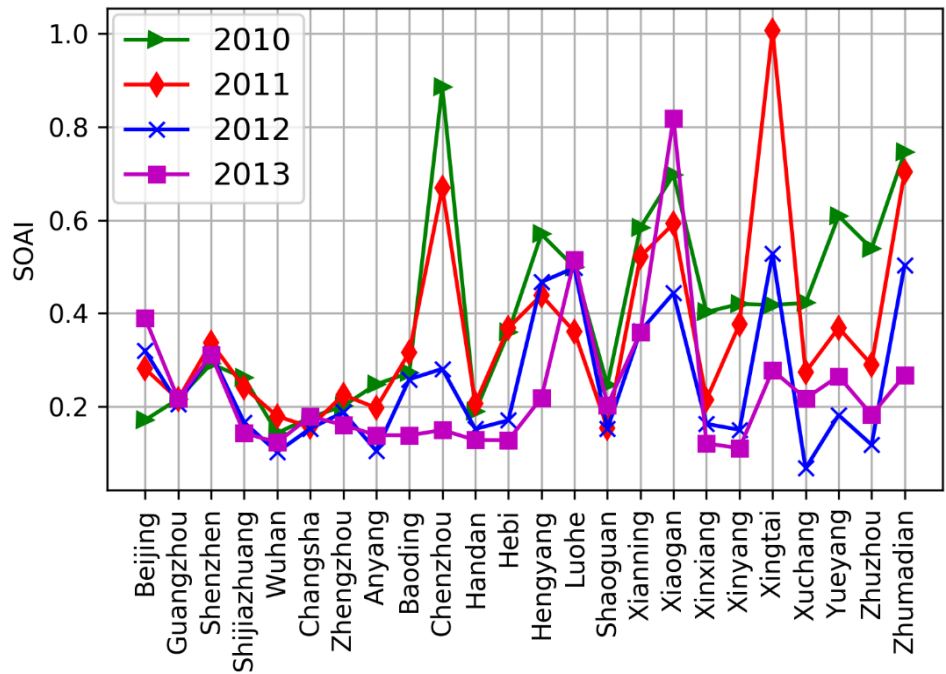

Figure 6(b). SOAI from 2010 to 2013

The cities in Figure 6 are sorted according to the grouping of the cities (FC, SC, TC) from left to right. To illustrate the differences between central cities and TC cities in SIAI and SOAI, the average level of central cities (FC cities and SC cities) and TC cities have been examined over the four years as shown in Table 8.

Table 8. Comparison of SIAI between the central cities and TC cities

|      | SIAI         |         | SOAI         |         |
|------|--------------|---------|--------------|---------|
| Year | Central city | TC city | Central city | TC city |
| 2010 | 0.1877       | 0.6317  | 0.3127       | 0.4787  |
| 2011 | 0.2694       | 0.7602  | 0.2834       | 0.4199  |
| 2012 | 0.242        | 0.5638  | 0.1791       | 0.2827  |
| 2013 | 0.3989       | 0.464   | 0.1475       | 0.2724  |

Table 8 illustrates that the SIAI and SOAI of the seven central cities are lower than those of the other cities. In other words, these seven central cities have less in-awareness and out-awareness than in other cities.

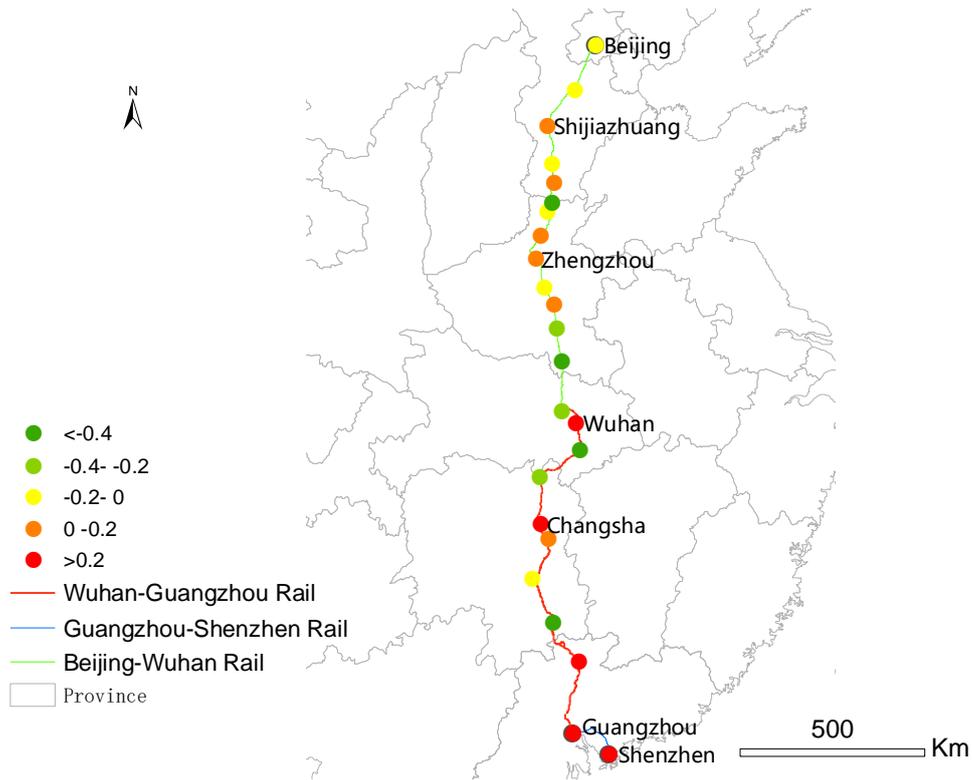

Figure 7(a). SIAI differences from 2013 to 2010

This paper considers the increase in SIAI and SOAI from 2010 to 2013 as shown in Figure 7 to examine the impact of HSR. Table 8 also shows that the average SIAI of six of the seven central cities is on an upward trend while that of 11 of the 17 TC cities is decreasing. In other words, HSR improves the degree of influence of the central cities.

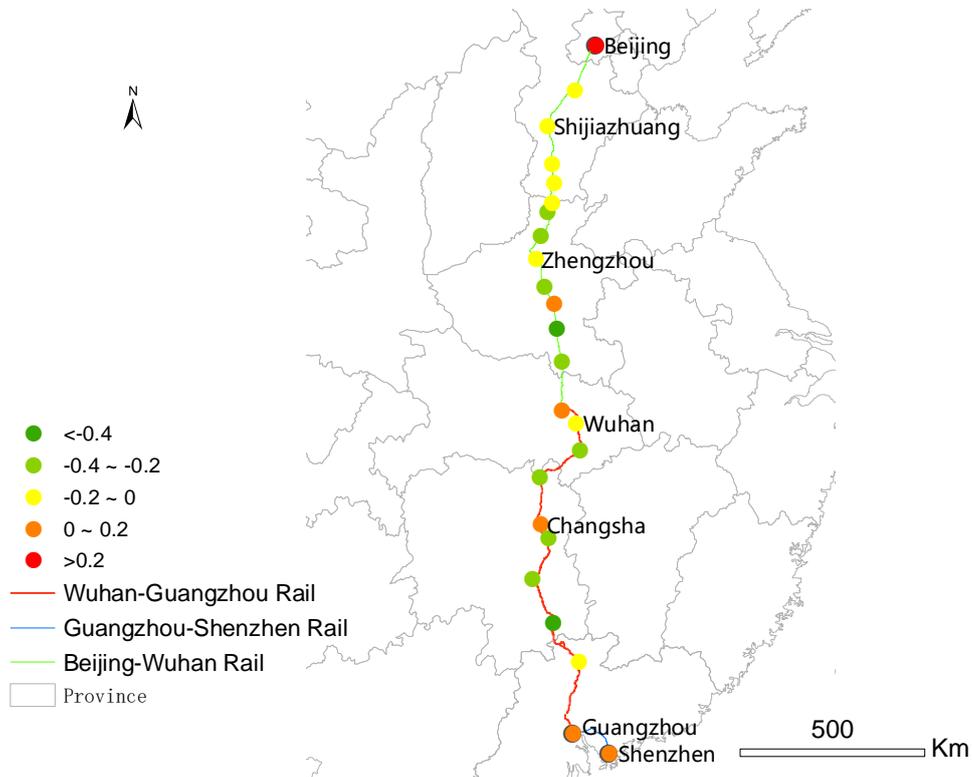

Figure 7(b). SOAI Differences between 2013 and 2010

The increment of out-awareness indices is depicted in Figure 7(b). The average out-awareness and the individual out-awareness of five of the seven central cities increased between 2010 to 2013, while the average out-awareness and the individual out-awareness for 15 of the 17 TC cities decreased. These findings indicate that HSR improves the spatial out-awareness of central cities in the TC cities.

**4.2.4. Impact and Travel Time**

The great advantage of HSR is that it reduces inter-city travel times and provides additional convenience. The reduction in time is an important factor that causes people to choose HSR as their mode of transport. Previous studies have verified that the significant change in inter-city relatedness is created by HSR in the two-hour travel time (Martín et al., 2014; Wang et al., 2015). To quantify the interaction between HSR impact and travel time, we adopt SIAI and SOAI to conduct the inter-city relatedness over cities from 2010 to 2013.

Nearly 100 trains are running through cities in the Beijing Shenzhen HSR. Even between the two cities, the trip times vary slightly due to stop conditions and changes in schedules. To simplify the calculation of the trip cost, trip time is set as the travel distance divided by the HSR average speed. The distance between cities can be evaluated using the spatial distance because the HSR is a relatively straight line. Results from the previous section show that HSR has opposing impacts on the seven central cities (FC & SC cities), as well as 17 non-central cities (TC). The spatial relatedness, SIAI and SOAI, will be examined by two groups separated based on whether or not the city is a central city. The experimental results are shown in Figure 8.

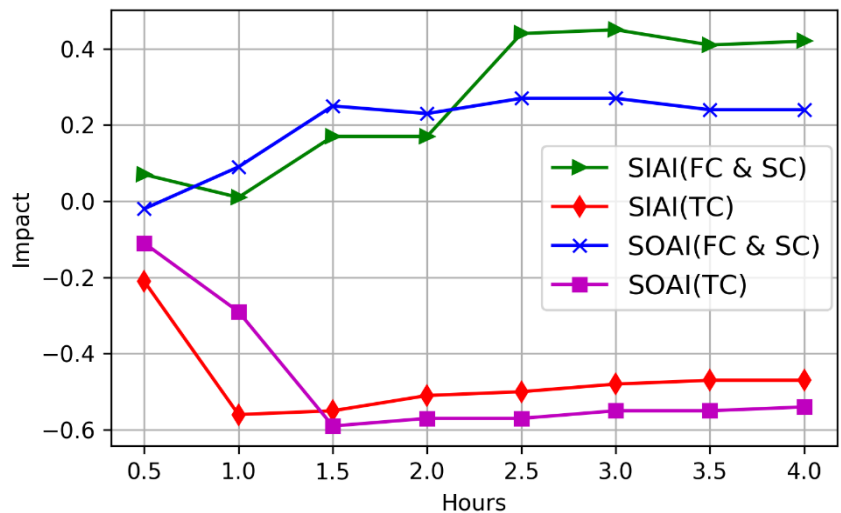

Figure 8. The improvement of SIAI and SOAI between 2010 to 2013 within given hours

On the local scale, there is a significant improvement in the spatial in-awareness and spatial out-awareness in the seven central cities since the HSR began operations. These trends are consistent with global trends. The SIAI curve reached the top at three hours, and slightly decreased after that. The SOAI curve peaked at 1.5 hours, and remains stable within the larger spatial scale. The strengthened SIAI and SOAI for central cities indicates that local relatedness for each is strengthened, and their centrality is enhanced.

The trend of SIAI and SOAI in non-central cities conflicts with the trends in the central cities. The SIAI and SOAI touch the bottom of the curve at one hour and 1.5 hours, respectively, and then begin to rise slowly. These data show that they have less influence and connection with the HSR in operation. It can be seen from Figure 9 that the impact of HSR in the two groups varies based on travel time. This shows the differences in radiation between the two groups. HSR improves radiation of the central city within a distance of three hours or more, while it decreases the non-central cities' radiation in the near distance (about 1-1hours). This change in radiation shows that the influence scope of HSR is not balanced, wherein the central cities' influence is larger than non-central cities' in spatial scope.

### 4.2.5. Channel effect

The channel effect demonstrates that there is a stronger connection between the starting city and the ending city than other cities, which can be quantified by the difference between the endpoint relatedness of the HSR and the total relatedness. There were two different tunnels formed between 2010 to 2013 in the HSR from Beijing to Shenzhen, due to the fact that the HSR started its service from Wuhan to Shenzhen at first and began its service for the rest of the line later. The first tunnel running from Wuhan to Shenzhen opened in 2012. The tunnel was discontinued when the second tunnel was constructed due to the HSR running from Beijing to Wuhan in 2013, and a new tunnel from Beijing to Shenzhen was formed. Both channel effect indices are examined from 2010 to 2013 as shown in Table 9.

where TE present the channel effect index derived the from perspective of economy.

At the year an HSR line begin running, the channel effects of the HSR line are shown with bold font.

Table. 9 TE and TS indices from 2010 to 2013,.

| Year | Wuhan To Shenzhen | | | | Beijing To Shenzhen | | | |
| --- | --- | --- | --- | --- | --- | --- | --- | --- |
| | $TE_{ij}$ | $TS_{ij}$ | $TS_{ji}$ | TS | $TE_{ij}$ | $TS_{ij}$ | $TS_{ji}$ | TS |
| 2010 | 0.0370 | 0.0503 | 0.3528 | 0.0880 | 0.0423 | 0.7087 | 0.4226 | 0.5295 |
| 2011 | 0.0384 | 0.0418 | 0.2686 | 0.0724 | 0.0422 | 0.7783 | 0.4939 | 0.6043 |
| 2012 | **0.0413** | **0.0429** | **0.3191** | **0.0757** | 0.0420 | 0.7831 | 0.5375 | 0.6375 |
| 2013 | 0.0414 | 0.0195 | 0.1767 | 0.0351 | **0.0408** | **0.8882** | **0.5980** | **0.7147** |

It can be seen from Table 9 that the TE index based on inter-city economy has some limitations when evaluating the channel effect: 1) The TE index does not reflect the inter-city channel effect correctly as the TE indices in bold show a lower-than-expected value, which is 0.0413 in this case. 2) The TE index need not indicate the impact of the HSR. For example, the TE index does not change significantly when a new HSR begins to run.

The TS index represents the inter-city influence caused by HSR, as a comparison point for the TE index. It shows a significant channel effect, as the TS indices are far greater than the expected value of 0.043 in both groups. In the first group, the trend of the TS indices in Table 9 decreased in 2010 and 2011, but increased in 2012 after the HSR began running. The index decreased again in 2013, at which time the Wuhan-Shenzhen was no longer a tunnel since Wuhan was no longer the endpoint of the HSR.

In the second group, the TS index increased significantly when the Beijing-Shenzhen tunnel formed. In addition, the channel effect is not symmetrical, especially in the first group where the index from Wuhan to Shenzhen (0.319) is far greater than the index for Shenzhen to Wuhan (0.043).

## 5. Discussion

Although HSR promotes economic growth (Chan and Yuan, 2017, Coronado et al., 2019), it also fosters an uneven relationship of economic, labor migration, and house values among cities (Gutierrez et al., 1996, Qin, 2017, Jin et al., 2017, Biggiero et al., 2017, Shao et al., 2017). How to identify the reason is very challenge because there are a lots factor may relate it.

The interaction between cities is a kind of available data can be employed to qualify the uneven relationships, and also can be used to observed the dynamic effect if this kind of data have temporal attribution. Thanks to the spatial information and high-precision time information contained in social media data, this paper develops a methodological framework to evaluate inter-city relatedness and HSR influence. The framework provides new metrics for dynamic examining the interactions between cities. Because the position and time of posts in social media is high precision, and the proposed framework could contribute to model impact of HSR in refined spatial and temporal scale.

The results show that the impact caused by HSR is unbalanced between central cities and non-central cities. The central cities have high in-awareness and low out-awareness, while non-central cities have lower in-awareness. And it is very interesting, the results also suggest non-central cities benefit more from HSR than central cities. This is in line with the idea of city policymakers trying to get high-speed rail through their own cities.

Meanwhile, a greatly promotes the spatial imbalance between central cities and non-central cities. HSR improves the spatial interaction of central cities and reduces the spatial interaction of non-central cities. The affected distance formed by HSR between the central city and the non-central city is different. The central city affects inter-city interactions of more than three hours of travel distance, while the non-central city has a greater impact within two hours of travel. Hence, policymakers in central and non-central cities should adopt different strategies when formulating policies.

In the field of management, the channel effect illustrates the phenomenon where majority shareholders stand in opposition for benefits for minority shareholders. Putting a high-speed rail line looking at a channel, the cities at either end of the high speed would be supposed to have a somewhat channel effect, especially the endpoint cities usually are large cities. In other words, more enhanced interaction between the two end cities is what meets expectations. Otherwise, it may be problematic and require more in-depth thinking on the part of policymakers.

This study is limited by the bias of social media content. On the one hand, since most social media users are young, they cannot yet be fully equated to the overall users of a city. On the other hand, it is possible that some cities have multiple name expressions or multiple cities have the same name. Therefore, there may be some noise in the basic interaction data obtained.

## 6. Conclusion

In this study we use social media data and develop a methodological framework to evaluate impacts of regional rail investment, HSR in China, on urban interaction. The framework measures inter-city in-awareness and out-awareness, and combines the spatial decay to define spatial in-awareness and spatial out-awareness. It also proposes a measurement for the channel effect. The relatedness model in this framework can not only evaluate the inter-city asymmetric interaction, but can also evaluate the inter-city relatedness more effectively than the traditional methodology. In addition, the relatedness model can be used to examine inter-city interaction. Furthermore, the framework sensitively reflects the changes of inter-city relatedness via a new perspective and provides an effective method to evaluate HSR impacts. Finally, the Beijing-Shenzhen HSR, which connects three of China's four first-tier cities, is used as a case study to demonstrate the applicability of the framework.

The follow-up research can incorporate more HSR lines to estimate the overall impact of HSR. In addition, migration data between cities can be combined with social media data to assess HSR's impact on inter-city traffic.